# Service Oriented Architecture in Enterprise Application

MS. Faathima Fayaza
Faculty of information technology
University of Moratuwa
Katubedda, Sri Lanka

*Abstraction-* **At present organizations try to achieve competitive advantages using the information technology (IT). Organizations not only use Information technology to manage their internal operations but also to collaborate with their customers and suppliers. For these organizations use enterprise applications. Also organizations expects from IT to address their shifting needs in on demand environment. So currently IT face challenges on integrating various system into function that can address organization on demand needs and span over the organizational boundaries. Now Service Oriented Architecture (SOA) imagine as architectural frame work address the issues on previous enterprise applications. This paper represents the Service Oriented Architecture in Enterprise Applications. The enterprise application gives a brief idea about enterprise applications. Afterward this paper addresses the main problems face by enterprise application on the evolution of enterprise applications and needs of service oriented architecture in enterprise applications. Afterwards discuss about Service Oriented Architecture and challenges on Service Oriented Architecture.**

## I.  INTRODUCTION

In the enterprise arena Service Oriented Architecture (SOA) now become as serious topic and more popular topic because in today's fast-moving global economy enterprise must be agile and flexible to encounter the shifting needs of process in on demand environment. SOA responds to forthcoming demand in enterprises are more quick and efficient. Service Oriented Architecture helps to enterprise to virtualized IT service and provides end-to-end enterprise integration.

Enterprise applications are big business highly complex applications which needs to satisfying hundreds or thousands of separate requirements. Enterprise applications are scalable, data-centric, distributed, mission-critical, user-friendly, and component-based. Enterprise applications may be deployed on a variety of platforms across corporate networks, intranets, or the Internet [8].

Service-Oriented Architecture is defined in various ways. One of them is "paradigm for organizing and utilizing distributed capabilities that may be under the control of different ownership domains" [16]. In simpler terms: SOA is a standardized set of methods for getting things done using whatever capabilities you have, wherever those capabilities reside, and in whatever fashion those capabilities can be organized or combined to deliver maximum benefit to the enterprise [16].

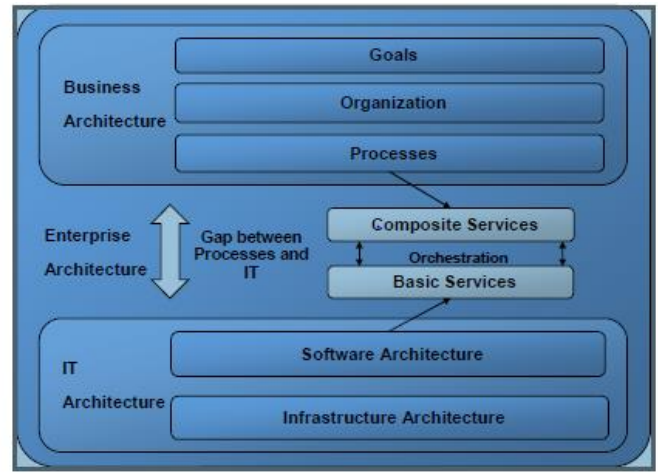

Figure 1 Service-oriented enterprise architecture [7]

Mostly entire enterprise is addressed by Enterprise Architecture. Enterprise Architecture comprising IT architecture and business. Also Enterprise Architecture addressing their interrelations. Business goals, organization and processes are in the business architecture. IT-architecture contains the software and infrastructure architecture. So enterprise architecture has to be responsive of all these different sub parts and their coherence [7].

Enterprise Information Systems are rapidly growing but previous Enterprise Application Integration technology cannot adapt change, because traditional enterprise application usually performs as independent application systems and focuses on point-to-point interconnection and hardly considers about other applications of the enterprise and the collaboration between two applications. So traditional Enterprise Applications are complicated because needs its own interface programs. And also it cannot provide necessary agility. Because of that SOA introduced and the main idea of SOA is providing agility by bridging the gap between the IT application layer and business process layer. I the past IT systems were designed as functional feed store but changing business operations often required completely different grouping of IT functions. So for this purpose service layer introduce. "By reuse of services bridging the business operations with IT is implemented in SOA. Therefore, SOA influences the service, the business process and the IT application layer"[1].

Today businesses face many complex issues like application integration, distributed software, multiple protocols, multiple platforms, the Internet, numerous devices, etc. So Enterprise applications are needs interoperable solutions. But before the Service-Oriented Architecture, applications didn't enable open interoperable solutions, but relied on proprietary APIs and required a high degree of coordination between groups. Service-Oriented Architecture is emerging as the premier integration and architecture framework in today's complex and heterogeneous computing environment and eliminates the problems of protocol and platforms. Service-Oriented Architecture delivers enterprise agility, by enabling rapid development and modification of the software that supports the business and help organizations to streamline processes so that they can do business more efficiently, and adapt to changing needs in an on demand and competition.

In earlier days programmers introduce modular design to reduce the rewriting code over and over. But when they try using modules into other applications they face problems. So developers proposed classes and object-oriented software to solve these problems. Once more, as software complexity grew, developers wanted a way to reuse and maintain functionality, not just code. So the component-based software introduced. But it is/was a good solution for reuse and maintenance, but it doesn't address all of the complexities developers faced. Today applications face many issues. These issues are can now eliminate by using Service oriented Architecture. [12].

Service-Oriented Architecture built applications based on services and loosely-coupled architecture. So they can be reused. A major benefit of Service-Oriented Architecture is that it delivers enterprise agility, by enabling rapid development and modification of the software that supports the business.

The remainder of this paper consists of six chapters. Chapter two describes about enterprise application. Chapter three describes main problems in enterprise applications. Chapter four describes needs of Service Oriented Architecture in enterprise applications. Chapter five describes about Service Oriented Architecture. And the chapter six describes the challenges in the implantation of service oriented architecture in enterprise applications. Finally, summarize and conclude the paper in chapter seven.

## II. ENTERPRISE APPLICATION

An enterprise application is a "big business application with set of integrated software modules and a central database that enables data to be shared by many different business operations and functional areas throughout the enterprise"[8]. In today's business world, enterprise applications are distributed, complex, scalable, component-based, and mission-critical. They are user-friendly, data-centric, and must meet stringent requirements for security, maintenance and administration. In short, they are highly complex systems [8].

Today an enterprise application performs many business functions such as processing, procurement, production scheduling, customer information management, and etc. And also enterprise systems provides many values to business such as, increase operational efficiency, provide firm-wide information to support decision making, enable rapid responses to customer requests for information or products, include analytical tools to evaluate overall organizational performance and etc. Enterprise application finally improves efficiency and output through the business level support functionality.

When developing an enterprise application needs to consider hundreds or thousands of separate requirements. Also when making development decision for a particular requirement it may affect many other requirements. So when developing enterprise application needs to have good understand to make it successful. Also enterprise application should more efficient and quick to response on demand environment. Enterprise information systems are rapidly growing so enterprise application must be adapt to the change. Over these enterprise application are specifically expect agility. But before the SOA applications not enable these behaviors. In enterprise application arena SOA emerging as the primer integration and architecture framework. SOA enable to reuse existing applications and delivers enterprise agility.

## III. EVOLUTION OF ENTERPRISE ARCHITECTURES AND PROBLEMS FACED IN THE EVOLUTION OF ENTERPRISE APPLICATIONS

Today enterprise applications are based on SOA concept. Enterprises use SOA to get agility and fast response in on demand. But before SOA developers use many concepts to develop the enterprise applications. In the history of enterprise application development developers introduce many concepts, each concept has developed for particular reasons and after that they did not be sufficient anymore, so they move to next evolution. Now developers mainly use the SOA to develop enterprise applications. So it's good to having looked on the history of enterprise application development and problems solved by each concept and problem solved by each concept.

*A. Monolithic Applications*

At the beginning efficiency was a major factor on development of monolithic systems and developed directly on purpose, Mostly system client and developers were either the same or had close relationship. So applications are strong coupling. Over time the monolithic systems needs to support for slowly changing process. By this every change in the system required the test of the whole system so maintenance effort was relatively high. In addition, every update made a system less structured and could be stared as another circle that was added. So the maintenance cost was become as a growing factor. Because of that the systems did not suffice anymore and developers need to consider about another systems [7].

*B. Component Based Architecture*

Related functionalities of an application are group together and components were created. Each component has an interface to communicate with other components. Component based system simplified the maintenance than monolithic system and functional expansion or replacement become easier than the one of a huge monolithic system. But component based system also still tightly coupled because components used proprietary interface so that it was still hard to integrate them with other components[7].

With the time, business processes need to integrate with different application. For that always implement a new interface with changed technologies, it's become serious cost problem. So it's leads to implement n different interfaces for n different technologies each of n components. Simply its use 1: n mechanism instead of n: n method. For these problems middleware technology is introduced. In middleware technology, each component offers an interface, which allows collaborating with all other applications connected to the middleware [7].

*C. Enterprise Application Integration (EAI)*

Main task of EAI was to integrate several applications with each other. This is possible by middleware technology like Common Object Request Broker Architecture (CORBA) or MQSeries which strongly decreased the integration effort in enterprise systems. Middleware simplify the integration of different applications. Also reduces the human efforts in retrieving the information. So it is saves a lot of effort and reduces the number of possible mistakes in a business process [7].

Applications, especially older ones, offer their functionality only via a graphical user interface but middleware cannot adapt to it automation was hindered once again. To overcome these kinds of problems GUI design had to be changed [7].

*D. Separation Of GUI*

Here GUI is separated and flexibly increased also all the functionality was reachable for automation purpose via middleware. Although its leads to increase design and implantation effort slightly. As a result GUI also uses middleware to connect application [7].

But business process lifecycle times were progressively decreasing and the flexible demands were increasing. So flexibility demands were still not satisfactorily fulfilled. This might be starting point of service-orientation [7].

Business experts expect the systems that have the alignment of business and IT. For this service concept is introduced. Service is a business function implemented by IT and it is not bound to a certain application [7].

Yahoo pipe is the example of GUI separation.

*E. Business- IT Alignment - Basic Services*

Here the Service Registry is introduced it keeps all the information about services that are required to use them. Service representation has advantages and disadvantages. Service can offer that span several applications, its leads to IT business alignment is the one of the advantage. But its increase the effort for identifying, implementing, and maintaining services as well as the related Service Registry [7]. But for the changing business processes GUI also need to change. For that using the portal and portlets technology web-based interfaces implemented. But up to this stage flexibility regarding the implementation of new process not reached its peak, because control of the process flow is distributed over humans and the whole application landscape. So new process application have to reprogram in their specific language. Because of these programmers has to find the code fragments that change the desired process control flow among other uninteresting code fragments.

*F. Hard-Wired Service Orchestration*

Basic services are representing simple business functions. Each service operation has to be invoked [7]. Orchestrating services is the process of automate and make them accessible in a relatively pure form. If it's done service it owns its call as hard-wired service orchestration. Data redundancies reduce by the orchestrations. A service orchestration, as done in the procurement service, easily offers the possibility to avoid redundant data and to retrieve data directly from other applications [7].

*G. Soft-Wired Service Orchestration*

Introduce new component of service oriented architecture, the Orchestration Engine [7]. In the hard-wired service orchestration, process control flow lies within service code. But in the soft-wired orchestration process control flow made explicit in an orchestration engine. It is able to interpret and execute special process models. The most common language for these process models is the Business Process Execution

Language (BPEL) [7]. It first brought up the idea of making process control flow more explicit and modeling it in an interpretable language. Now, in combination with the middleware used it is possible to link process activities to realized services. At the same time, the business process oriented tailoring of services eases the modeling of executable process models [7].

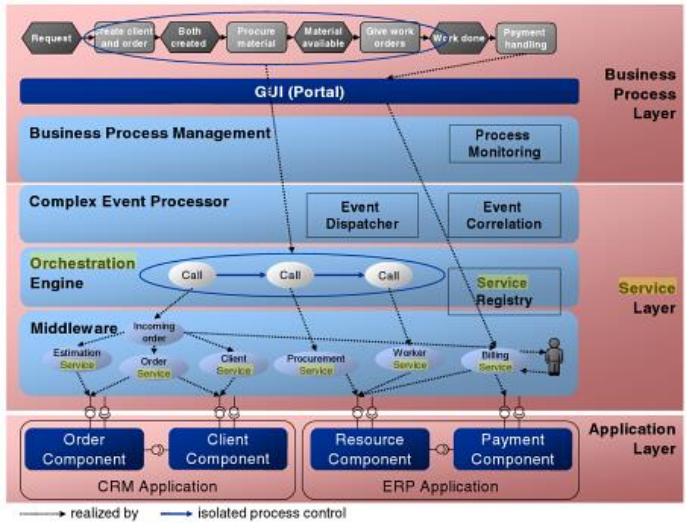

Figure 2 Fully developed service oriented enterprise architecture [14].

## IV. LIMITATIONS NEEDS OF SERVICE ORIENTED ARCHITECTURE IN ENTERPRISE APPLICATIONS

SOA have been introducing to enterprise for many reasons. SOA have been used by specialists to solve the complex problems arising from evolutionarily grown, heterogeneous application landscapes in large business.

Enterprise needs are rapidly growing but traditional enterprise application systems cannot adapt change, because traditional enterprise application usually performs as independent application systems and focuses on point-to-point interconnection and rarely considers other applications of the entire enterprise and the cooperation between two applications. So traditional Enterprise Applications are complicated because needs its own interface programs. And also it cannot provide necessary agility [1].

Reuse and cost cutting are the main reasons for introduction of enterprise application integration (EAI). But using EAI flexibility could not be meeting. But considering the increasing complexity of an application landscape following the introduction of a SOA, the re-use and cost cutting arguments will lead to disappointment. SOA offer a great potential to increase corporate agility [6]. Using the previous experience on enterprise application development participants introduce SOA leads to better adaptation of corporate information systems to change business processes and to better business process support in general.

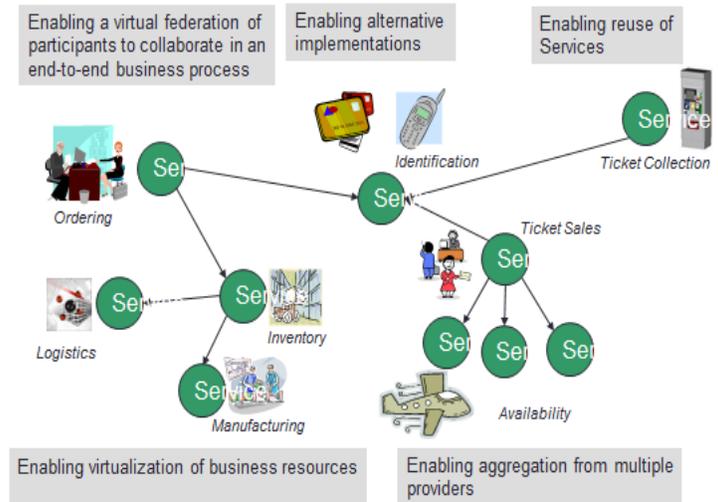

Figure 3 SOA to enable flexible, federated business processes

Although agility and flexibility are often used as synonyms but flexibility as one part of agility only it can understand from the discussion in production management [6]. E.g. Yusuf et al. understand agility as the capability to adapt to unexpected changes, whereas flexibility is focused on expected changes only [6]. Flexibility is "built in" by considering in early design stage but it cannot to contribute to unexpected changes, because only expected changes could be considered at the design stage. Yusuf et al. define agility as "the successful exploration of competitive bases (speed, flexibility, innovation proactivity, quality and profitability) through the integration of reconfigurable resources and best practices in a knowledge-rich environment to provide customer-driven products and services in a fast changing market environment" [6]. Agility contributes to unexpected changes. Using agility of SOA enterprise having many advantages some of them are:

- Feedback at different level
- More efficient development processes
- Adequate business infrastructure
- Cost saving
- Risk management
- Evolution approach
- Independency from technology
- Reuse

Service-oriented architecture (SOA) is a method to build applications for distributed systems based on components. Main advantages of SOA are the reusability, flexibility and complication.

It does connect various systems in a unified and general manner using services through the well-defined interface and agreements among these services. Services are independent of the hardware platform, operating system, and programming language. At present, most enterprises use Remote Procedure Call (RPC) for interaction with each other. Common RPC technologies include: COBRA, Distributed Component Object Model (DCOM), Remote Method Invocation (RMI), and Web service. COBRA and DCOM are less open than RMI and Web service in terms of supporting protocols, data format, and interfaces, so is not suitable to be used to publicize the related technologies and to realize SOA system[1].

SOA provides an excellent platform for enterprise applications. With the help of Web service, it can realize the integration of enterprise applications conveniently, safely, and efficiently [1].

It sends application functions to end users or other services as services. As an object/component deployed on Web, SOA is a self-describing, self-contained, and modular application. Its most distinguished feature is that it can cross multiple platforms. It supports multiple programming languages, runtime networks and environments, and can be explained with the XML language.

## V. SERVICE ORIENTED ARCHITECTURE

Nowadays enterprise expects flexibility to meet their shifting needs of business process in an on demand environment. Hence Enterprise applications are desires interoperable results. however before the Service-Oriented Architecture, applications didn't allow open interoperable resolutions, on the other hand, applications depended on proprietary APIs and requisited a high degree of coordination between applications. Today enterprise use Service Oriented Architecture to get the virtualized IT service and end-to-end enterprise integration. [1]

Service-Oriented Architecture defined in various ways, "Service-oriented architecture is a Client/Server design approach in which an application consists of software services and software service consumers (also known as clients or service requesters). SOA differs from the more general client/server model in its definitive emphasis on loose coupling between software components, and in its use of separately standing interfaces" (Gartner).

In today's heterogeneous and complex computing environment Service-Oriented Architecture is emerging as the leading integration and Architecture framework and eliminates the difficulties on various devices, application integration, distributed software, varying platforms, and varying protocol. Also Service-Oriented Architecture enable flexible, Combined Business Processes, and business process optimization and the Real Time Enterprise (RTE)

The architectural concepts accompanying with SOA are not new it's already developed on CORBA, DCOM and others but key assurance is agility.[8] Agility is the most importance delivers of Service Oriented Architecture, by allowing modification and rapid development of programs delivers the agility to business processes. [3]

SOA is method of design, development and management of both application and the software infrastructure where: all software is organized into business services that are network accessible and executable, service interface are based on public standards for interoperability.

Service-Oriented Architecture is an architectural approach that allows distributed deployment by expose enterprise data and business logic as loosely coupled, discoverable, structured, standards-based, coarse-grained, stateless units of functionality called services. Furthermore allows reusability by choose a services provider and access to existing resources exposed as services. By allowing reusing the existing applications Service-Oriented Architecture enables enterprise to influence existing investments. Another importance is composability by allowing assemble new processes from existing services that are exposed at a desired granularity through well defined, published and standards complaint interface. Also provide interoperability by share capabilities and reuse shared services across a network irrespective of underlying protocols or implementation technology.

Key characteristic of SOA: quality of service- response time, security and performance, service is cataloged and discoverable, data are cataloged and discoverable, protocols use only industry standards

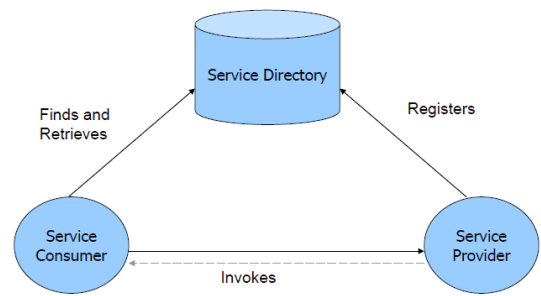

Figure 4 Service Oriented Architecture Model [7]

A SOA has three major parts; service provider, service consumer, and service directory. Service providers are the parties who build service and make available service. Service consumers are the clients who consume services. Service directory is the place where service providers register the services and consumer search for services. Service directory provide following services:

1. Scalability of services; can add services incrementally.

2. Decouples consumers from providers.

3. Allows for hot updates of services.

4. Provides a look-up service for consumers.

5. Allows consumers to choose between providers at runtime rather than hard-coding a single provider.

*A. Service*

Building block of SOA is autonomous services. Service is a reusable component. Service are not homogeneous they are divers they can include old systems and new systems. Service comes into the data and can change the data according to business need from one state to another. Services remains stable but its configuration will be changed. A service is the way how date is accessed. But all the services are not a web service.

A service exposed it functionality using three properties,

- Interface agreement to service is platform-independent

- Can dynamically located and invoked

- Service has its own states, so autonomous

When model service it's essential distinguish consistent interface, need to be public standard. Client from any language, any Operating System (OS) and anywhere can get the service by the interface agreement on platform independents. Service providers are published their service on service registry. Services are available on service registry. Clients can find the services by look-up mechanism from the service registry. According to the client need client can select the services.

In a service life cycle have 3 stages.

Expose – how service is implemented

Compose- created services are combined into business process

Consume –make available to end user

Service has boundaries. Boundary denotes the border between interface and implementation. WSDL used to publish the boundary.

*B. Messages*

Services are interacting through the messages. On service interface agreement is defined the message return and accept. Service provider and customers interact through the massage so massage structure is importance. Messages are building using XML because it does provide scalability, all the functionality and granularity request by message.

## VI. CHALLENGES IN THE IMPLANTATION OF SERVICE ORIENTED ARCHITECTURE IN ENTERPRISE APPLICATIONS

Today enterprise applications used Service Oriented Architecture to meet they are shifting needs. But when implementing Service Oriented Architecture need to face many challenges. For some challenges there are some ongoing researches and for some challenges some researchers suggested some results by researches.

When implementing the Service-Oriented Architecture Service identification is the very first challenge. Today same business function is provides by multiple service providers. So when selecting the service needs to answer the question like what business functionality needed and what is provided by service? [10]

Service location is another challenge. Services operate on business entities, occupant within system records. Ideal location of service execution is system record. In distributed architecture business data spread across multiple applications. So service location is challenge. Service domain definition, service packaging, service orchestration, service routing service governance and service messaging standards adoption also some challenges of Service Oriented Architecture.[10]

"There are many researches going on functional layers of SOA. In the basic service layer service definitions addressing, functional, non-functional aspects associated with services are related problems and they address by many researches" [17].

"Services are in addition to its specific function support for sets of protocols and formats addressing extra-functional concerns such as transaction processing and reliable messaging. Transactional coordination in service-oriented computing is address by Tai et al"[17]. "The authors claim for the use of declarative policy statements to advertise and match support for different transaction styles (direct transaction processing, queued transaction processing, and compensation-based transaction processing) and introduce the concept of and system support for transaction coupling modes as the policy-based contracts guiding transactional business process execution" [17].

At the development time SOA requires service description in (Universal, Description, Discovery, integration) UDDI repository system by using this client can develop program that can bind to and interact with service of specific type. On that understanding the execution semantic is a weight task [17]. For this quality of service management framework based on user expectations suggested by Deora et al [17].This framework collects expectations as well as ratings from the users of a service and then the quality of the service is calculated only at the time a request for the service is made and only by using the ratings that have similar expectations [17].

SOA UDDI (Universal, Description, Discovery, integration) delivers business-category browsing mechanism for review and select services to developers. It works based on keyword-search might be upgraded by introducing more powerful matching approaches. "By combining syntactic and semantic comparison algorithms of Web Services Description Language (WSDL) document hybrid matching approach suggested."[17]. "In a peer-to-peer based framework is examined that allows advertising and finding services using keyword-based search, ontology-based search and behavior-based search in a highly decentralized and dynamic environment. In addition, the framework provides mechanisms so that users may express and query the quality of services".[17]

## VII. CONCLUSIONS

Today's enterprise application arena, SOA is the most powerful and popular topic due to it agility and reaction to an upcoming demand in enterprises are more efficient and quick. Enterprise applications are big business highly complex applications which needs to satisfying hundreds or thousands of separate requirements. Also enterprise applications are need to adapt to the changes driven by enterprise very quick and fast. From the last decades evolution on the enterprise architecture style SOA is implemented. While on evolves there are many methods have to be adapted and on each method they satisfy some need and afterward the aspect enterprise increase specialists need to going for another method through the evolution developers now use SOA to develop the enterprise applications. Method are adept during the evolution are monolithic applications, component based architecture, enterprise application integration (EAI), separation of GUI and business- IT alignment - basic services. Services are published in the service registry. Service can orchestrate in two ways. It can be hard-wired service orchestration (incoming order service), so services use directly other services. Or it can be soft –wired, by the orchestrate engine allows to the process model with the services and that can execute these process plans afterward [14].

Today almost all the businesses are intent in the enterprise application to full fill their needs in an on demand and to get competitive advantages. And businesses expect to work together not only in the organization but also with in the supplier and customer. To full fill these needs today most of the enterprise applications are based on the Service-Oriented Architecture. Because main problem in the previous applications are didn't enable open interoperable solution, but relied on proprietary APIs and required a high degree of coordination between groups. But Service-Oriented Architecture provides answer for these kinds of issues in a cost effective manner.

Main idea of SOA is bridging the gap between the between the business process layer and the IT application layer. In the past IT systems were designed as functional silos but changing business processes often required completely different grouping of IT functions. The service layer is introduced for this purpose. By reuse of services bridging the business processes with IT is implemented in SOA. Therefore, SOA influences the business process, the service and the IT application layer.

While developing the SOA developers face many challenges. For these challenges there are many research going on also some challenges addressed by some researches.

However Service-Oriented Architecture built applications based on services and loosely-coupled architecture designed to meet the business needs of the organization so they can be reused. A major benefit of Service Oriented Architecture is that it delivers enterprise agility, by enabling rapid development and modification of the software that supports the business processes.